\documentclass[usenatbib]{mn2e}
\usepackage{graphicx}
\usepackage{array}
\usepackage{multirow}
\usepackage[font=small,format=plain,labelfont=bf,up,textfont=normal,up]{caption}
\usepackage{float}
\usepackage{txfonts}

\title[Spectropolarimetric observations of Mrk 6]{Variability in Spectropolarimetric properties of Sy 1.5 galaxy Mrk 6}

\author[Afanasiev et al.]{V. L. Afanasiev$^{1}$\thanks{E-mail: vafan@sao.ru}, L. \v C. Popovi\'c$^{2,3,4}$,
 A.I. Shapovalova$^{1}$, N.V. Borisov$^{1}$,  D. Ili\'c$^{3,4}$\\
$^1$ Special Astrophysical Observatory of the Russian
Nizhnij Arkhyz, Karachaevo-Cherkesia 369167, Russia \\
$^2$ Astronomical Observatory, Volgina 7, 11060 Belgrade 74, Serbia \\
$^3$ Department of Astronomy, Faculty of Mathematics, University
of Belgrade, Studentski trg 16, 11000 Belgrade, Serbia\\
$^4$ Isaac Newton Institute of Chile, Yugoslavia Branch
 }

\begin{document}

\maketitle

\label{firstpage}

\begin{abstract}
Here we present an analysis of
spectro-polarimetric observations of type 1.5 AGN Mrk 6, performed  with 6m telescope SAO RAN in
12 epochs (2010 -- 2013). Additionally, the inter-stellar mater (ISM)
polarization has been observed and its contribution to the AGN spectral polarization is taken into account.

We measured Stokes parameters and determined the polarization parameters in
12 spectra with and without correction for the  ISM polarization. We estimated the
time lag  between the unpolarized and polarized continuum flux
variation {  of} about $\sim 2$ days,  that indicates a compact scattering region {  which}
 contributes to the
 polarized continuum variability.  The polarization in H$\alpha$  is complex, showing
three prominent components in the BLR, one redshifted around +3000 kms$^{-1}$ that corresponds to the red shoulder in
H$\alpha$, and
two blue-shifted around -2000 kms$^{-1}$ and -6000 kms$^{-1}$.

We found that the ISM polarization has a very significant influence on the
measured AGN polarization parameters. After {  correcting} the observations
 for the ISM polarization we were able to detect the Keplerian motion in the BLR.

We give a new method for the black hole mass estimation using spectro-polarimetric observation in the line profile, finding
the black hole mass in Mrk 6 of
$M_{BH}\sim 1.53\cdot 10^8M_\odot$, that is in a good agreement with reverberation estimates.

\end{abstract}


\section{Introduction}

Spectropolarimetry is a powerful tool for
probing the innermost part of Active Galactic Nuclei (AGNs), since
polarized light provides information about the
radiation that is coming from the center {  of an AGN} and from {  the scattering region}
 assumed to be between the central source and observer.
Therefore, Spectropolarimetric observations have been very important in studying the AGN inner
structure, especially after the discovery of the hidden polarized broad-line emission in a Seyfert
2  {  galaxy} NGC 1068 \citep{am95}.

This discovery has proved that AGNs type 1 and 2
are intrinsically the same type of objects, but viewed from different orientations,
i.e. in type 2 galaxies, the direct view of the active nucleus is blocked by
the optically and geometrically thick torus of molecular gas and dust \citep{an93}.
Subsequently, many other type 2 AGNs have been observed to have polarized broad lines
\citep[see e.g.][etc.]{mg90,tr92,yo96,he97,tr01,kis02,ki02,lu04}

On the other hand, for broad-line AGNs such as Seyfert 1 galaxies, an  analysis of the
polarization structure across the broad line profiles can potentially give information on the
geometry of the broad-line region (BLR). Moreover, there is a possibility to investigate
the polarization mechanisms across its velocity field \citep{ma98}.

It is shown by \cite{gm94} that the polarization of broad lines is roughly similar to the
continuum polarization, {  supporting} the idea that the polarization has been due to scattering. The investigations
of the properties of the polarized light in the AGN with broad lines \citep{sm02,sm04,sm05,gg07}
have shown that the broad line emission originates in a
rotating disc and it is scattered by material in either the equatorial plane of
the disc or orthogonal to it, the so-called equatorial and polar scattering
models \citep[see][and references therein]{sm02,sm04,sm05,gg07}\footnote{ Note here, that in some case
of type 1 AGNs, the polarization of the broad lines almost completely vanishes \citep[see e.g.][]{kis08}}.
Additionally, the synchrotron radiation
also may contribute to the optical polarization in {radio-loud} type 1 AGNs similar as in the case of blazars
\citep[see e.g.][]{gs93}.

The variability in the polarized light is of a particular interest, since a
time-dependent variation of the polarized broad line flux and {  profile}  as well as correlation between
the polarized and unpolarized continuum and broad lines can give more information about the innermost structure of
type 1 AGNs. Actually, variability studies such as reverberation
mapping might not be completely convincing, since they only consider the total
line and continuum flux, but not potentially valuable information from polarized light.
Note here that a convincing and self-consistent BLR model has to
account for the broad-line {  profile variability} and {  the} variability in
both flux and polarization, and continuum-line responses in both polarized and unpolarized light
\citep[see e.g][]{ga09}.
In this context \cite{ma98} analyzed high S/N ratio polarized spectra of NGC 4151 observed in two epochs,
and found {  correlations}  between the scattering
axes of the different parts of the broad H$\alpha$ line profile and the morphological
axes of the host galaxy. {  The author}  noted that a full three-dimensional model
of the BLR of NGC 4151 will require higher S/N ratio spectropolarimetry with more frequent
time sampling. Also, recently \cite{ga12} investigated long-term variability in the polarized
continuum of NGC 4151 and found  that the polarized flux of NGC 4151 appears to lag the unpolarized
flux by $\sim$8 days which is comparable to the light-crossing
time of the BLR and that the dust in the torus is ruled
out as the source of polarization in the continuum.

{  We present the}  spectropolarimetric observations (monitoring) of Mrk 6 in
{  more}  than two year period, from 2010 to 2013 (12 spectra) in the optical
spectral range (covering the H$\alpha$ and H$\beta$ wavelength region). Mrk 6 (IC 450) is a Seyfert 1.5 galaxy (z=0.0185,
 m(B)=14.29, M(B)=-20.41) that has been
observed in the frame of the Spectropolarimetric monitoring campaign of AGN
with the 6-m BTA Telescope \citep{af11}
Mrk 6 is a good candidate for studying continuum and line polarization structure variation, since it is a bright
object with {  the} significant variability in
broad emission lines and with the BLR that  most-likely has a complex geometry, structure and kinematics
\citep{Ro94, E93, sm02, DS03}. Previous optical monitoring has shown that the continuum and broad
Balmer lines are strongly variable \citep{E93, Se99, DS03, Do12}, and the strong blue asymmetry is
seen in Balmer-emission lines \citep{E93, Kh11, Do12}. The analysis of the line profiles
variability and time-lags gives that the BLR kinematics in Mrk 6 is a
combination of Keplerian (disc-like) and infall gas motion \citep{Do12}.
Also, some signature of the outflowing dense hydrogen gas-cloud has been
seen in Balmer-emission lines \citep{Kh11}. This could be supported by
the radio measurements as radio-jets {  have}  been observed in this object \citep{Ca95}.
Recent VLA observations show spectacular radio structures \citep{Kh06}
on kpc-scales, that cannot be explained by super-winds ejected by a nuclear starburst,
but most preferably by an episodically powered precessing jet
that changes its orientation \citep{Kh06}. On the other hand, the X-ray observations
favor the warm absorption model, where the absorbers are probably originating
from the accretion disc wind \citep{Sc06, Mi11}. The Spectropolarimetric analysis
of Mrk 6 shows that the broad line emission originates in a rotating disc, while
the polarized emission can be described with the scattering in the equatorial
plane \citep{sm02,sm04}.

The polarization in the continuum and broad lines can be caused by different mechanisms, first of
all {  there is} the intrinsic polarization caused by the geometry of the emitting region\footnote{ as e.g. 
disc like geometry of the BLR
will have a small percent of polarization in difference with the case of a spherical BLR where we expect
unpolarized light}, after that light can
be scattered in the polar and equatorial plane \citep{sm05}.
These {  scatterings}  can be caused by a jet or { outflowing}  gas, as well as {  by} some gas in
equatorial plane between the BLR and torus, and finally torus can be a source of
 {  scattering}.

In this paper  we analyze  the  spectrophotometric observations of Mrk 6 obtained in 12 epochs
in order to constrain the continuum and  BLR structure and nature of the scattering region.
 Additionally we discuss the observed variability in polarization in the frame of the unified model and
complex AGN structure. The paper is organized as following: in \S 2 we describe observations and
data reduction, in \S 3 the results of our analysis are given, in \S 4 we discuss results, and {  in \S 5 we} outline our conclusions.

\begin{figure}
\centering
\includegraphics[width=8cm]{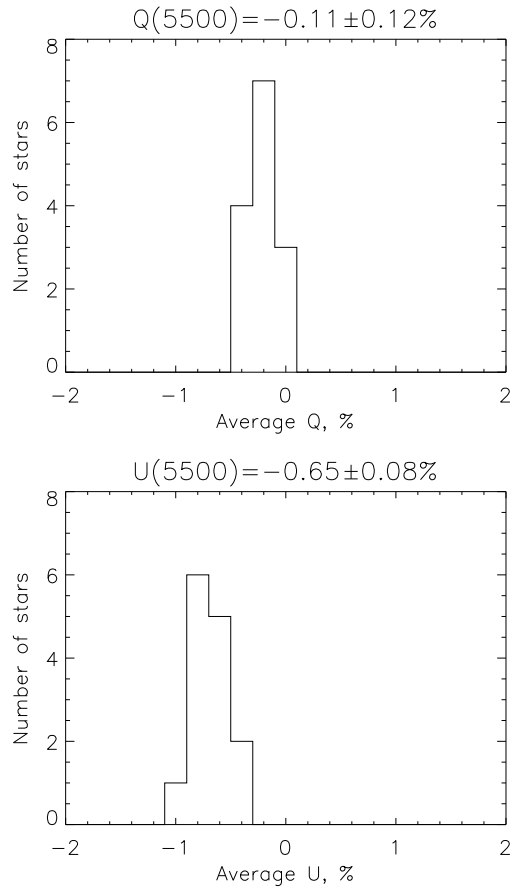}
\caption{Histograms of ISM polarization parameters measured from 14 stars around Mrk 6.} \label{fig1}
\end{figure}

\section{Observations and data reduction}

\subsection{Observations}

The  instruments and method of observation have been described in
more details in \cite{aa12}, and here will not be repeated, only we give
 some basic and specific details of observations and data reductions.

Spectropolarimetric observations of Mrk 6 have been performed with 6m telescope of SAO RAN using the modified
spectrograph SCORPIO \citep[see in more details in][]{am05,am11} in the mode of spectro-polarimetry and polarimetry.
{  The Wollaston prism has been used as
 an analyzer},  which provides that, at the same time,  the  light beam  from the source has been
registered in two long-slit spectra (with a slit of  1-2$^{\prime\prime}$ width and  120$^{\prime\prime}$ height)
which are perpendicular {  to} each other
in  the polarization direction. Behind  the slit, in the focal plane, the rotation super-achromatic plate with
$\lambda/2$ -- phase has been installed \citep[more detailed description of the instrument can be found in][]{aa12}.

We observed the spectra of Mrk 6 AGN and standard stars with the position of the analyzer in four different angles:
0, 45, 22.5 and 67.5 degrees. The polarization of standard stars (taken from \cite{hs82} and \cite{sc92}) was observed
each night for two purposes: i) to control  the instrumental polarization, and ii) to calibrate  the polarized
AGN spectra. Additionally, we observed non-polarized stars as spectro-photometric standards.
We tested the instrumental polarization, and found that it is less than 0.05\% .

The differences between our measurements of polarization standards and these in the catalogue
 were 0.1-0.2\% for polarization and 2-4 degrees for the polarization angle.

In Table \ref{tab1} the log of observations  is given. The date of observations, Julian date,
total {  exposure time}, number of cycles, seeing, slit, mode and
spectral resolution are listed in Table \ref{tab1}. We observed in three different modes, using gratings
 VPHG940 and VPHG1200 which
covered the H$\beta$ and H$\alpha$ wavelength ranges. Additionally we have observations of the ISM polarization
with the filter V with  $\lambda$(max) 5500\AA.
The number of cycles denotes a number of observations for each position angle in the phase plate or
Polaroid, i.e. one cycle corresponds to
the observations in all four above mentioned angles. The spectral resolutions given in Table {  \ref{tab1}} were
estimated using the FWHM of the lines from  the night-sky in the integral (AGN+night sky) spectra.

\begin{table*}
\begin{center}
\caption[]{Log of observations.}\label{tab1}
\begin{tabular}{lllccccc}
\hline \hline
Date obs.&JD&Total exp.&Num. &Seeing&Slit&Mode&Spectral \\
&240000+&sec.&of cycles&arcsec&arcsec&&resolution (\AA)\\
\hline
2010.11.05&	55505&	2400&	5&	2.5&	2&	1&	7.5\\
2011.08.27&	55800&	3360&	6&	2&	1&	1&	5.4\\
2011.11.21&	55886&	2400&	5&	1.2&	1&	1&	5.2\\
2012.02.14&	55971&	3600&	5&	 4&	2&	1&	7.8\\
2012.04.21&	56038&	2880&	6&	1.5&	2&	1&	7.9\\
2012.05.18&	56065&	2400&	5&	1.5 &	2&	1&	8.0\\
2012.08.25&	56164&	3360&	7&	1.1&	1.5&	1&	6.5\\
2012.09.11&	56182&	2400&	10&	1.5&	2&	2&	6.6\\
2012.09.14&	56184&	2400&	10&	1.5&	2&	1&	7.5\\
2012.10.08&	56208&	3600&	10&	2.5&	1&	2&	6.6\\
2012.11.12&	56243&	3360&	7&	1.1&	2&	1&	7.8\\
2013.02.06&	56328&	900&	10&	1.5&	Image&	3&	-\\
2013.02.08&	56330&	2400&	10&	1.5&	2&	1&	8.1\\
\hline
\end{tabular}
{\\
Mode of observations:\\
1 --	Wollaston$+$phase plate $\lambda$/2$+$grating VPHG1200, spectral coverage  3700-7300 \AA\\
2 --	Wollaston$+$phase plate $\lambda$/2$+$grating VPHG940,  spectral coverage 3700-8400 \AA\\
3 --	Polaroid$+$filter V, $\lambda$(max) 5500\AA,  FWHM 850 \AA\\
 }
\end{center}
\end{table*}

\subsubsection{Polarization of the inter-stellar matter (ISM)}

The observed linear polarization of an object is a vector composition of the ISM polarization
 and polarization of the object (in this case the Mrk 6 AGN), i.e. $\vec{P}_{obs}=\vec{P}_{AGN}+\vec{P}_{ISM}$.
The ISM polarization, as it is well known, is depending on the Galactic latitude and it has a strong changes in the
rate of polarization and in the polarization angle  on one degree scale on celestial sphere. That is
connected with non homogeneous distribution of the ISM. In a number of papers,
the ISM polarization has been taken into account   as
a function of the Galactic extinction E(B-V)  for different latitudes as it is described in \cite{se75}.
However,
the problem with this method is that it does not take into account the direction of the ISM polarization vector,
 i.e. vector
$\vec{P}_{ISM}$ has direction and intensity and both quantities should be taken into account \citep[see e.g.][]{kis04}.
Therefore, here we take into account the ISM polarization vector by measuring polarization of a number
of stars around the AGN, that in principle represents the ISM polarization. For these observations we
used the wide-field  photometry (dichroic polaroid in the field with diameter of 6$^\prime$),
since within the range of 0.45-0.8 mkm  a typical change in the ISM polarization
is smaller than 10\% from the maximum,
 i.e. for Galactic longitudes  $>25^\circ$ the estimate of  wide-field $\vec{P}_{ISM}$ is satisfactory
 for this purposes.

For the ISM polarization estimate we used  dichroic polaroid with three fixed angles -60,  0  and 60 degrees,
and with the 6$^\prime$  field. We
considered  only bright surrounding stars {  within}  3$^\prime$ around Mrk 6  in the V filter.
We found 14 {  brightest} stars in the field and for each of them calculated parameters
Q and U.\footnote{for detail procedure of polarization parameters calculation see \cite{aa12}}.

In Fig. \ref{fig1} we give histograms of
 our measurements. The estimates of the averaged polarization parameters of the ISM are:
$Q_{ISM} = -0.11 \pm 0.12\%,\  \   U_{ISM}   = -0.65 \pm 0.08 \%, \ \  P_{ISM}   = 0.66 \pm 0.14 \%,\ \
 \varphi_{ISM}  = 130 \pm 4.1^\circ$.

Note here that in the catalogue of \cite{he00} the rate of ISM polarization is
 changing around 0.4-1.7\% and angle of ISM
polarization around 70-180$^\circ$ in the radius of 10$^\circ$, that is not in contradiction with  our estimates.

\subsubsection{Data reduction}

The data reduction includes standard procedure for long-slit spectroscopy, bias, flat field, geometrical
correction along the slit, correction of the spectral line curvature, the night-sky subtraction, spectral sensitivity of the instrument
 and spectral wavelength calibration.
 For the absolute flux calibration we used the fluxes of  the [OIII]5007  and [SII]6717/31 spectral lines,
taking the fluxes from \cite{Do12} (for [OIII]5007  6.9$\pm$0.11$\times \rm 10^{-13} erg\ cm^{-2}s^{-1}$,  and for
 [SII]6717/31 1.61$\pm$0.09$\times \rm 10^{-13} erg\ cm^{-2}s^{-1}$).
From the long-slit spectra of the galaxy we extracted the spectra of the AGN. One can expect
the contribution of the stellar radiation from the AGN Mrk 6 host galaxy, but our estimate of this contribution shows
that the host galaxy is depolarizing the AGN continuum radiation (around $<1$\%).

\begin{figure}
\centering
\includegraphics[width=8.5cm]{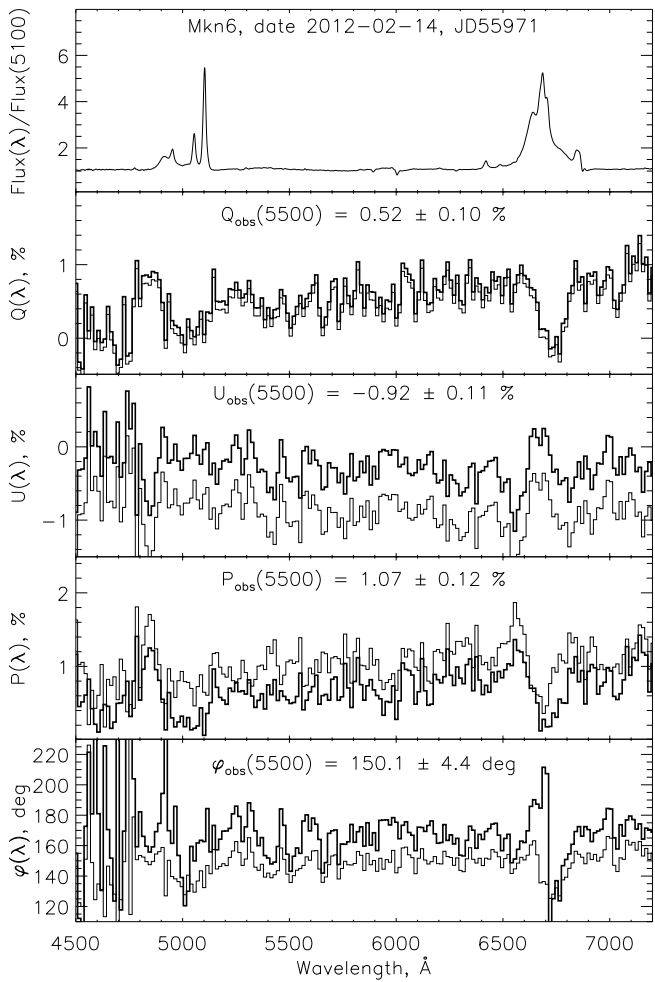}
\caption{The spectrum and its polarization parameters  observed on Feb 14, 2012.
From top to bottom: the observed spectrum of Mrk 6, parameters Q and U, and corresponding {  percentage}
 (p) and angle ($\varphi$) of polarization. The thin line represents observed and thick line represents corrected
spectra for the ISM polarization.} \label{fig2}
\end{figure}

We integrated spectra along the slit, since the procedure of decomposition (of the observed light as a function of
wavelength along the slit) of
the host galaxy and AGN  increases the statistical errors. For the estimates of
the Stokes parameters ($I_\lambda,\ Q_\lambda,$ and $U_\lambda$),  and consequently
the  linear polarization {  percentage}  ($P_\lambda$) and polarization angle ($\varphi_\lambda$), we calculated for each spectral channel
$$F_\lambda={I_o(\lambda)-I_e(\lambda)\over{I_o(\lambda)+I_e(\lambda)}},$$
where $I_o$ is the ordinary, and $I_e$ the extraordinary intensity measured  in different angles of
$\lambda/2$-phase plate (0, 45, 22.5 and 67.5 degrees).
After that we found the Stokes parameters as:
$$Q_\lambda={1\over 2}(F_\lambda(0^\circ)-F_\lambda(45^\circ))$$
and
$$U_\lambda={1\over 2}(F_\lambda(22.5^\circ)-F_\lambda(67.5^\circ))$$
Then we calculated $p$ and $\varphi$ as:
$$P_\lambda=\sqrt{Q_\lambda^2+U_\lambda^2} \ \ \ \ \varphi_\lambda={1\over 2}{\rm arctg}(U_\lambda/Q_\lambda)$$

The method of the calculation of polarization parameters is,
in more details, given in \cite{aa12}. Note here that in the reduction we did not take into account cosmic rays and
different types of smoothing or using the optimal aperture photometry, since all of these algorithms
(procedures) can influence
on the appearance of an artificial instrumental polarization.  In further analysis, the polarization parameters
($I_\lambda, Q_\lambda, U_\lambda,  P_\lambda$ and $\varphi_\lambda$) have been {  robustly}  estimated as average in a spectral window of
 $25 - 30$ \AA\ for all cycles of measurements, the number of the windows for  different observation data are
between 5 and 10.
 This type of measurements gives a good estimate of statistical errors of the measured parameters.
 An example of  the Mrk 6 polarized spectrum is present in Fig. \ref{fig2}, where the measurements of Stokes parameters,
the percentage of the linear polarization and the angle of the polarization are shown.
In Fig. \ref{fig2} thin lines denote observed spectra without {  the} correction {  of}
the ISM polarization, and thick lines the data after {  the} correction {  of}
 the ISM polarization. In Table \ref{tab2} we give polarization parameters
with and without the correction {  of the}  ISM polarization and
further in the text we will use corrected polarization parameters.

{ As one can see from Table \ref{tab2}, the ISM polarization
can significantly affect the polarization parameters, especially in the case where the  polarization rate is not high (as it is
usual for type 1 AGNs). The difference between the rate of polarization in the spectrum with and without correction on the 
ISM polarization can be almost 100\% and difference in the polarization angle around 40-50 degrees. 
We should mention here that other methods can be used for the ISM polarization  correction, as e.g.
a correction of the ISM polarization using  $E_{B-V}$ \citep[see e.g.][]{kis02} where the contribution ISM polarization 
have to be compared in the QU diagram for expected values ​​of the ISM polarization (defined by $E_{B-V}$). However,
we performed a direct measurement of the ISM polarization from the stars around
Mkn 6, that allows us a simple vector subtraction for the ISM polarization correction.} 

\begin{figure}
\centering
\includegraphics[width=8cm]{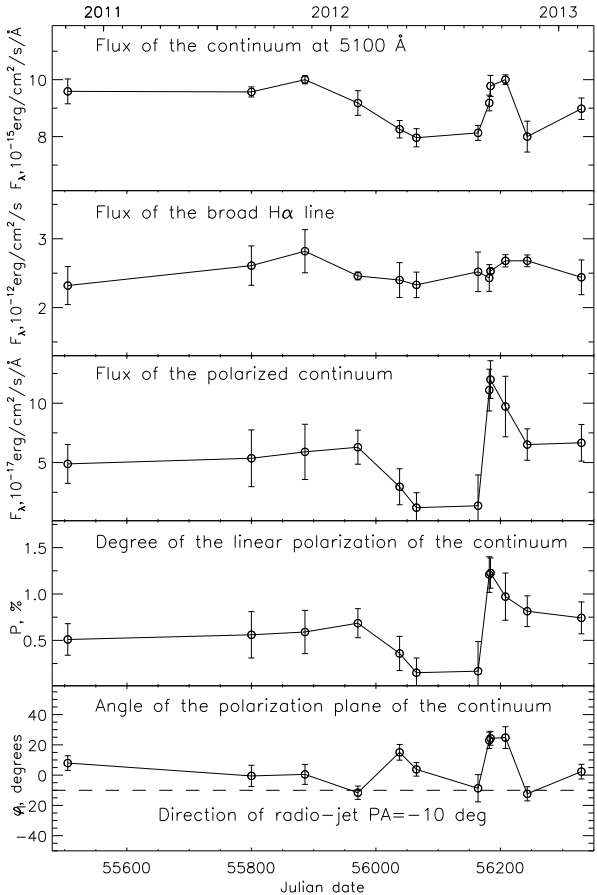}
\caption{Variation of the  Mrk 6 spectral and polarization parameters in the observed period.
From top to bottom the curves of: the continuum flux at 5100 \AA\ (rest wavelength), the H$\alpha$ line flux, the polarized continuum flux,
{  the percentage}  of polarization in the continuum, and $\varphi$ the angle of polarization. The horizontal dashed
line on {  the} 5th panel denotes the jet direction.} \label{fig3}
\end{figure}

\begin{table*}
\begin{center}
\caption[]{Observed continuum and H$\alpha$ fluxes, stokes parameters, {  percentage} of polarization and polarization angle of the continuum.
With 'corr' are denoted the values of the {  percentage} of polarization and polarization angle
which are corrected {  for the}  ISM polarization.}\label{tab2}
\begin{tabular}{lcccccccc}

\hline \hline
\\
JD&Flux(5100)&Flux H$\alpha$&$Q(V)_{obs}$&$U(V)_{obs}$&$P(V)_{obs}$&$\varphi(V)_{obs}$&$P(V)_{corr}$&$\varphi(V)_{corr}$\\
240000+&&&&&&&&180$^\circ+$\\

\hline
55505&	9.59$\pm$0.44&	2.32$\pm$0.28&	0.38$\pm$0.13&	-0.51$\pm$0.11&	0.67$\pm$0.17&	153.3$\pm$4.8&	0.51$\pm$0.17&	     8.0$\pm$4.8\\
55800&	9.57$\pm$0.18&	2.61$\pm$0.29&	0.45$\pm$0.20&	-0.66$\pm$0.15&	0.95$\pm$0.15&	150.5$\pm$4.3&	0.56$\pm$0.25&	   -0.5$\pm$7.1\\
55886&	10.01$\pm$0.15&	2.82$\pm$0.31&	0.48$\pm$0.30&	-0.64$\pm$0.31&	0.98$\pm$0.41&	149.7$\pm$8.6&	0.59$\pm$0.23&	    0.5$\pm$8.6\\
55971&	9.18$\pm$0.43&	2.46$\pm$0.06&	0.52$\pm$0.10&	-0.92$\pm$0.11&	1.07$\pm$0.12&	 150.1$\pm$4.4	&0.68$\pm$0.16	&   -10.5$\pm$4.4\\
56038&	8.26$\pm$0.31&	2.40$\pm$0.25&	0.20$\pm$0.13&	-0.47$\pm$0.13&	0.53$\pm$0.18&	147.8$\pm$5.2&	0.36$\pm$0.18	&   15.1$\pm$5.2\\
56065&	7.96$\pm$0.32&	2.33$\pm$0.19&	0.04$\pm$0.09&	-0.63$\pm$0.13&	0.65$\pm$0.16&	136.9$\pm$4.5&	0.15$\pm$0.16&	    3.8$\pm$4.5\\
56164&	8.13$\pm$0.26&	2.52$\pm$0.29&	0.05$\pm$0.21&	-0.70$\pm$0.24&	0.76$\pm$0.32&	137.3$\pm$9.0&	0.17$\pm$0.32&	   -8.7$\pm$9.0\\
56182&	9.18$\pm$0.28&	2.43$\pm$0.19&	0.73$\pm$0.15&	0.22$\pm$0.12&	0.78$\pm$0.19&	188.5$\pm$5.4&	1.21$\pm$0.19&	  23.0$\pm$5.4\\
56184&	9.78$\pm$0.37&	2.53$\pm$0.04&	0.70$\pm$0.12&	0.27$\pm$0.11&	0.76$\pm$0.16&	190.5$\pm$.6&	1.23$\pm$0.16&	  24.3$\pm$4.6\\
56208&	9.99$\pm$0.17&	2.68$\pm$0.09&	0.52$\pm$0.19&	0.09$\pm$0.17&	0.56$\pm$0.25&	185.6$\pm$7.2&	0.97$\pm$0.25&	  24.8$\pm$7.2\\
56243&	8.01$\pm$0.54&	2.69$\pm$0.08&	0.63$\pm$0.15&	-0.99$\pm$0.07&	1.20$\pm$0.23&	151.4$\pm$6.4&	0.81$\pm$0.17&	-12.3$\pm$4.7\\
56330&	8.98$\pm$0.38&	2.44$\pm$0.25&	0.63$\pm$0.14&	-0.59$\pm$0.10&	0.88$\pm$0.17&	158.3$\pm$4.9&	0.74$\pm$0.17&	   2.3$\pm$4.9\\
\hline
\end{tabular}
\\
{UNITS: Flux(5100) in	$10^{-15}\rm erg\ cm^{-2}s^{-1}\AA^{-1}$, 	Flux H$\alpha$ in
$10^{-12}\rm erg\ cm^{-2}s^{-1}\AA^{-1}$; 	$Q, U$ and $P$ in percents; $\varphi$ in	degrees\\}
\end{center}
\end{table*}
\begin{figure}
\centering
\includegraphics[width=8cm]{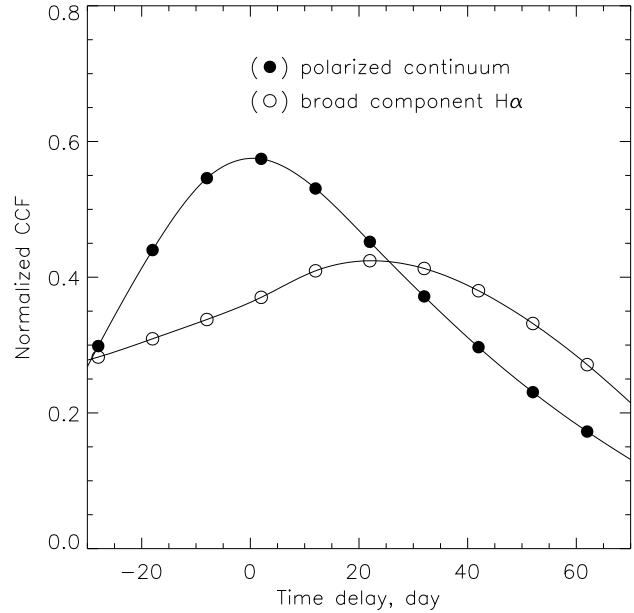}
\caption{Cross correlation functions between the unpolarized and polarized continuum flux (full circles)
and with the continuum and H$\alpha$ line (open circles).} \label{fig4}
\end{figure}

\begin{figure}
\centering
\includegraphics[width=8cm]{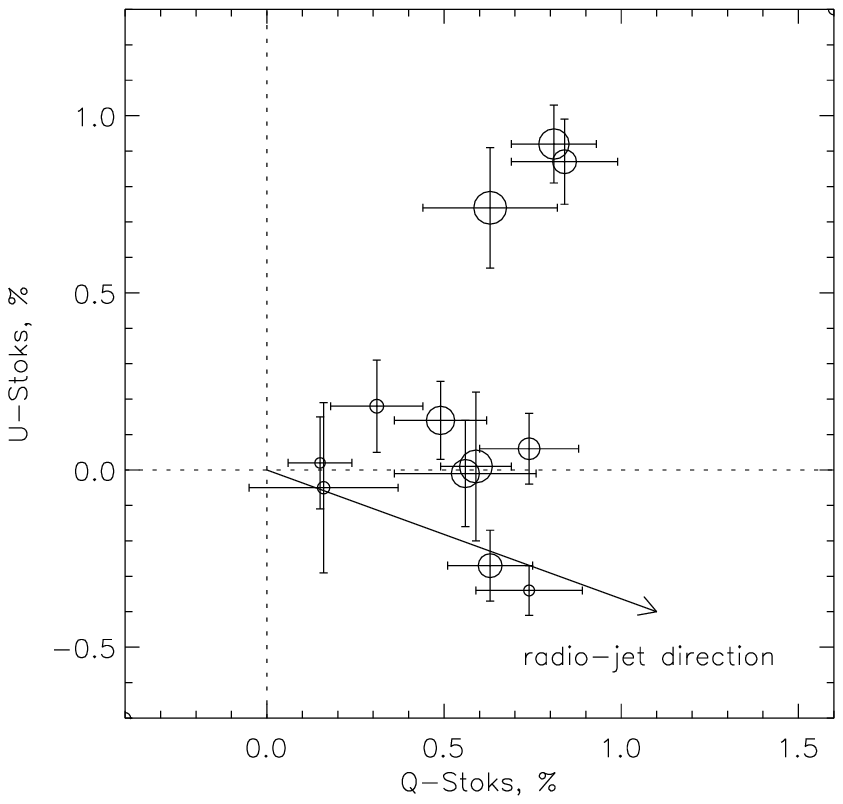}
\caption{Variability of the continuum in the Stoke parameter (UQ) space. The dimension of the
circles corresponds to the
intensity of the continuum flux at 5100 \AA . The { arrow} represents the position { and direction}  of the radio jet
in the UQ space.} \label{fig5}
\end{figure}

\section{Results}

The observed spectra of Mrk 6 covered the broad H$\alpha$ and H$\beta$ emission lines.
The broad lines showing an asymmetric profiles and
strong narrow emission lines (see Fig \ref{fig2}, first panel) are present in the spectra.
  As one can see from  Fig. \ref{fig2} the $S/N$ in
 the Stokes parameters
 is significantly  higher in  H$\alpha$ than in H$\beta$, therefore we measured polarization data for H$\alpha$.
In Table \ref{tab2} we give the measured flux (in the line and in the continuum at 5100 \AA\ rest wavelength)
and corresponding polarization parameters --
 the observed  ($P(V)_{obs}$ and $\varphi(V)_{obs}$),
 and  corrected for the
ISM polarization ($P(V)_{corr}$ and $\varphi(V)_{corr}$). As we noted above,
 the ISM polarization has a significant influence on the measured AGN polarization parameters.

\subsection{Polarization in the continuum}

The continuum of Mrk 6 is polarized on the level of $\sim$0.5\%
(see Fig. \ref{fig2} and Table \ref{tab2}), i.e. during the monitoring period the continuum polarization was changing between
0.15\% and 1\%, and angle $\varphi$ between $\sim$170$^\circ$ to $\sim$200$^\circ$.
The variability of the continuum, H$\alpha$ line, the polarized continuum and the polarization parameters
from JD2455505 (Nov. 2010) to JD2456330 (Feb. 2013) are present in Fig. \ref{fig3}.
As it can be seen from  Fig. \ref{fig3} the variation in the polarized
continuum nearly follows the variation in the unpolarized one.
The continuum at 5100 \AA\ (rest wavelength) {  changed for}  about 1.5 times, while
the line flux was changing with a smaller amplitude than the
continuum one. It seems that {  the} polarized continuum flux has a higher amplitude of the variation that the unpolarized one.
There is a high change in 2012 in both, the  polarization (4th panel in Fig.	\ref{fig3}) and in the polarization angle
(5th panel in Fig. \ref{fig3}). The polarization angle of the continuum is nearly following the jet orientation,
 however in the active phase the angle is changing {  for} around 20-30 degrees.

To estimate the dimension of the scattering region we cross correlated the unpolarized continuum with polarized one.
For this purpose we used a very simple method,
first we interpolated the light curves in the unpolarized and polarized continuum
(as well as in the H$\alpha$ line), after that we use the {  cross correlation function} (CCF) and found a lag
between the unpolarized and polarized continuum of
1.6$^{+0.1}_{-0.8}$  days. Also, we cross correlated the unpolarized
continuum and H$\alpha$ (see Fig. \ref{fig4}) and found
a lag of 20.3$^{+2.29}_{-2.11}$  days that is in a good agreement with the estimate given
 by \cite{Do12}. They found the H$\alpha$ lag of 27 days.
 { Note here that the time lags we measured for the polarized continuum and the BLR are shorter than the overall time sampling
 of the light curves, that is a general problem in the reverberation method \citep[see e.g.][]{sh13}, but generally the CCF method gives
 more-less reasonable lag values and as we can see in Fig. \ref{fig4} there is a sharp peak in the CCF between of the
 non-polarized and polarized
 continuum lag. Moreover, our estimate for the H$\alpha$ lag is in a good agreement with the reverberation estimate given by \cite{Do12},
 that is in favor of the used method. }

Such a small lag between the unpolarized and polarized continuum rules out scattering in
the torus of Mrk 6, it seems that the scattering region is not so far from the BLR,
and that it is more compact than the BLR.
Probably the scattering has some connection with the jet, or/and outflow { that is nearly to the jet direction}.
This confirms Fig. \ref{fig5}, where in the
UQ space the measured
points are presented with also the projection of the jet (the jet angle is taken from \citep{Ca95}
which projection on the UQ space is {  presented} with the { arrow} in Fig. \ref{fig5})\footnote{
Note here that the jet direction in Fig. \ref{fig5} corresponds to angle $2\cdot PA_{jet}$
(i.e. 340$^\circ$), since we take that in the $UQ$ plane the angle of the jet projection is $PA_{jet}={1\over2}{\rm arctg}(U/Q)$}.
The measured points are very close to the projected jet direction in the
UQ space, but during outbursts, there is
a drift that is {  nearly} perpendicular to the jet direction in the UQ space.
The size of circles corresponds to the
intensity of the continuum (bigger circles -- higher continuum flux), and it {seems} that
 {  the largest}  differences between polarization parameters and jet projection are in the active AGN phase (see Fig. \ref{fig5}),

\begin{figure}
\centering
\includegraphics[width=8cm]{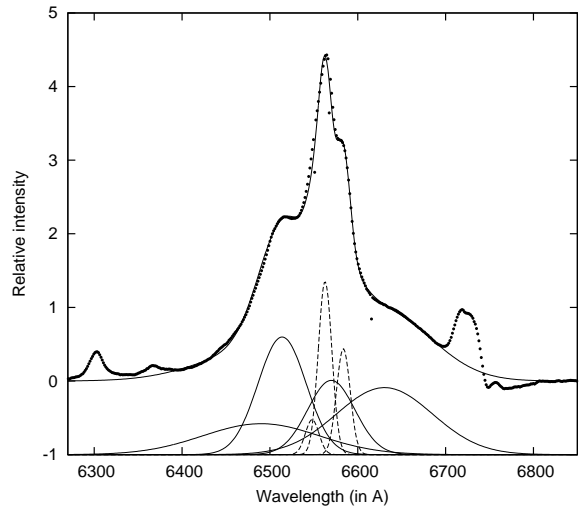}
\caption{Decomposition of the H$\alpha$ line profile into  Gaussian components.
{  Dashed lines denote the narrow lines { (the central, the most intensive is the narrow H$\alpha$,
and two from left and right sides are the [NII]6583, 6548 lines)}, and solid lines the broad line components}.} \label{fig6}
\end{figure}

\begin{figure}
\centering
\includegraphics[width=8cm]{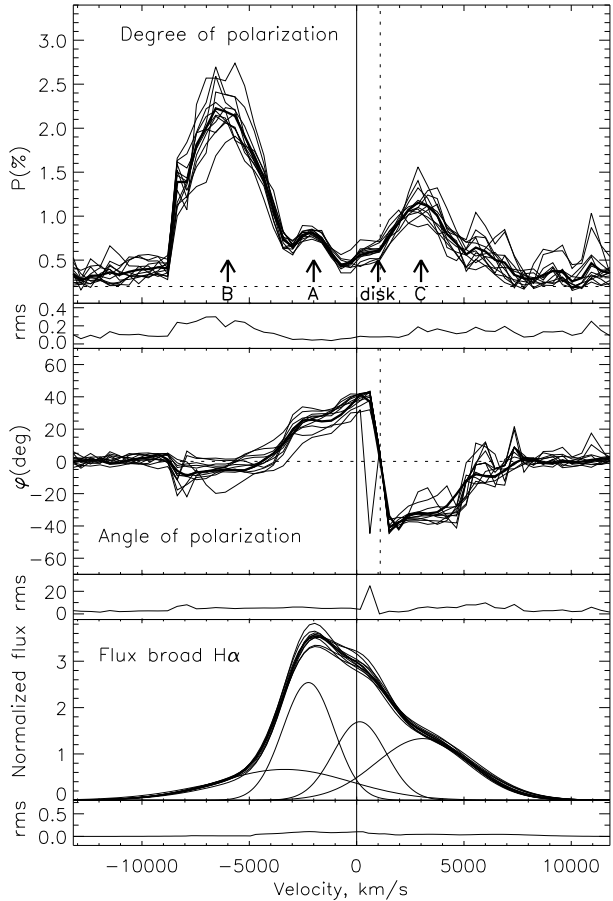}
\caption{Polarization in the H$\alpha$ line, from top to bottom: the rate of polarization,
the polarization angle, and
the broad H$\alpha$ line profile decomposed into four broad components. {  Thin lines denote}
 the observations from ten epochs, and {  thick lines denote}  the averaged value of parameters.
Below each plot the rms is {  given}.} \label{fig7}
\end{figure}

\subsection{Polarization of the BLR light}

As it can be seen in Fig. \ref{fig3}, the total flux of H$\alpha$ varied with a smaller amplitude than
the continuum, and it seems that there is no big {  change}  in the line profile in more than the 2-years period.

In order to investigate the polarization of the BLR light, we investigated the  polarization and
 polarization angle across only the broad
H$\alpha$ profile, since, as it was mentioned above, the S/N is higher in H$\alpha$ than in H$\beta$.
An inspection of the polarized Mrk 6 spectra shows that from 12 H$\alpha$ spectra,
two of them are with a lower {S/N}, therefore in the further analysis of the broad
line polarization  we use ten H$\alpha$ spectra with a good resolution.

The first step in our analysis was to subtract all narrow lines and obtain
only the broad component of  H$\alpha$. This was done using
the standard techniques of the multi-Gaussian decomposition \citep[for details see e.g.][]{po04}.
The decomposed H$\alpha$ profile is shown in Fig. \ref{fig6}. As it can be
seen in Fig. \ref{fig6} the line profile is very complex and can be fitted
with four broad Gaussian functions. The broad line profile has a
big red asymmetry (see Fig.\ref{fig7} - bottom), but the intensive blue peak
(located at $\sim-$2000 ${\rm km\ s^{-1}}$ is prominent in the line profile (see
Figs. \ref{fig6} and \ref{fig7}).

{ The observed shape of the polarization and polarization angle as a function of wavelengths in the H$\alpha$ line profile 
depend on the geometry and kinematics in the BLR. One can  expect that the  polarization across the line profile is caused by
the equatorial scattering in the inner part  of a dusty torus \citep[see e.g.][]{sm05}. 
As it was mentioned above, the polarization of the continuum emission 
is probably caused by scatter in the inner part of the BLR (in the accretion disk and/or jet). 
Therefore, to consider the polarization  only in the broad line one
should subtract the observed continuum polarization. Therefore,
to find polarization in the  broad line we performed following procedure: First  we take into account the ISM polarization 
in Stokes parameters $Q$
and $U$, after that { we multiplied $Q$ and
$U$ by total line flux}; Finally, 
from these parameters ($Q\times I$ and $U\times I$) we subtracted the polarization
parameters of the continuum.

The polarization and polarization angle across the line profile are shown
 in Fig. \ref{fig7}.
The measurements of the broad line polarized flux, especially in the line wings 
\citep[where $P/\sigma_P<$ 0.7, see][]{ss85}, is subject to bias.
In order to avoid the bias problem in the degree of the broad line polarization for small values of polarized flux in the line,
we assumed that for the far wings ($<-9000\ \rm kms^{-1}$ and $>+9000\ \rm kms^{-1}$) the line flux is on the zero-scale,
therefore the degree of the polarization and polarization angle for this, extreme, velocities are on the zero-scale. For rest
of the measurements (in the velocity interval from -9000 $\rm kms^{-1}$ to +9000 $\rm kms^{-1}$)
we calculated $rms$ as $\sigma_P=\sqrt{\sigma_Q^2+\sigma_U^2}$ and $\sigma_\varphi(deg)=28.2\cdot \sigma_P$ and plot in Fig. 
\ref{fig7}. As it can
be seen in Fig. \ref{fig7} the $p/rms_p$ is enough large in the considered velocity interval
that cannot be significantly affected by bias.}

It is interesting that polarization has three peaks, {one around -6000 ${\rm km\ s^{-1}}$ (bump B),
second around -2000 ${\rm km\ s^{-1}}$ (bump A) and third
around +3000 ${\rm km\ s^{-1}}$ (bump C in Fig.\ref{fig7}). The dashed line in Fig. \ref{fig7} corresponds to the zero scale
of polarization angle ($\varphi\sim 0^\circ$) and the disc component, that shows smaller rate of polarization.}
 Such a complex polarization shape in the line profile
 indicates a complex picture of polarization of the BLR light.
 {The redshifted (+3000 kms$^{-1}$) corresponds to the red shoulder in the H$\alpha$ wing, and
the blue-shifted (-2000 kms$^{-1}$) to  the blue peak in the broad H$\alpha$  line profile
(see Fig. \ref{fig7}). These two components seems to be with the nearly same offset from
the zero-scale of the polarization angle (dashed
line in Fig. \ref{fig7}). The component with the peak at -6000 kms$^{-1}$ cannot be seen in the line profile, and may
indicate outflowing gas in the central part of the BLR.}
The most interesting is the shape of $\varphi$ as a function of the velocity; the shape nearly follows one
expected in the BLR with the predominant Keplerian motion and zero point is { shifted to the red (+1000 ${\rm km\ s^{-1}}$)}.

\begin{figure}
\centering
\includegraphics[width=8cm]{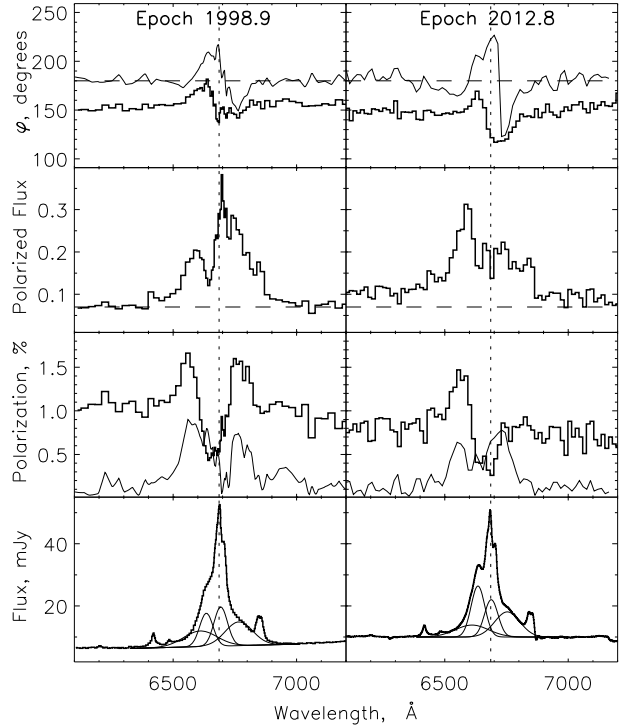}
\caption{Comparison of the observation of Mrk 6 H$\alpha$ given by Smith et al. (2002) - left panel,
and  our observation - right panel. In both spectra
we showed the  observations with (thick line) and without (thin line) the correction {  of}   the ISM polarization.
From the top to the bottom: The polarization angle, polarized line flux, percentage of polarization and the observed H$\alpha$
line profile decomposed into  Gaussian components.} \label{fig8}
\end{figure}

In Fig. \ref{fig7} we presented also rms of p, $\varphi$ and normalized line profile\footnote{The broad lines are
normalized  to the corresponding total line flux}, and there is no big
{  change}  in the {  profile} of H$\alpha$ line as well as in the shape of polarization parameters.

To explore variability in a longer period, we compared our observations (corrected on ISM polarization) with ones given
in \cite{sm02} (seen  Fig.\ref{fig8})\footnote{Note here that we digitized the \cite{sm02} data and applied
the same procedure for ISM polarization corrections as in our data, see \S 2}.
As it can be seen in Fig. \ref{fig8}
 there are significant changes in the H$\alpha$ line profile, as well as in the
polarization parameters. The broad line profile observed in 1999 also can  be fitted with four broad components,
but  the blue peak in the line profile
is more intensive in our than in observations of \cite{sm02}. The polarization in the red part observed in 2012 is significantly
smaller than in the 1999 observations as well as zero velocity depolarization.
It is interesting that in previous observation the $\varphi$ of the continuum is
 practically the same as in our one, and also the function $\varphi$ vs. velocity stays nearly the same as our, also indicating
a Keplerian motion in the BLR \citep[see][]{sm05}.

\begin{figure}
\centering
\includegraphics[width=8cm]{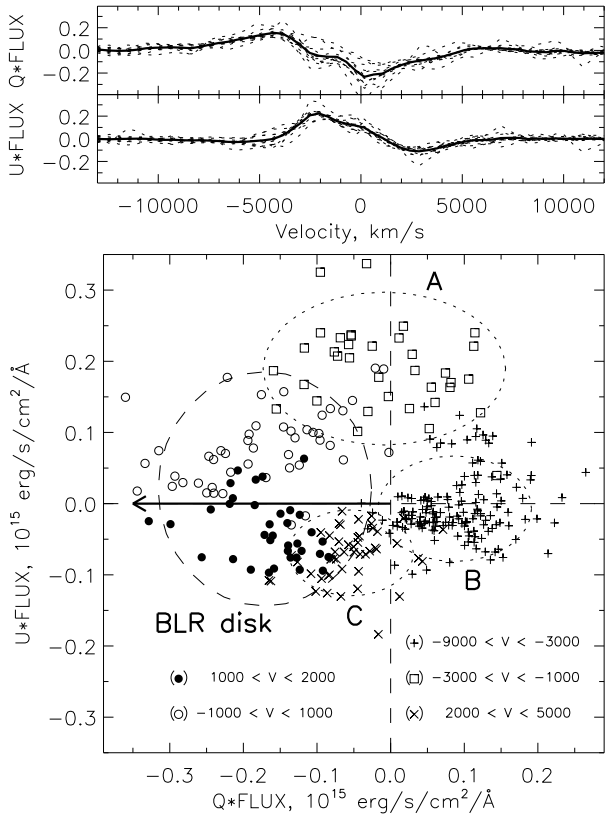}
\caption{Averaged Q$_\lambda$  and U$_\lambda$ parameters (upper panel) and the projection of velocities along
the H$\alpha$
profile on the UQ space (bottom panel). { $-$Q is transformed in the radio-jet direction { (the arrow on the plot)}.
The notations of  different velocities are given on the plot. }}
\label{fig9}
\end{figure}

\subsubsection{The complex BLR or/and scattering region}

As it can be seen in Fig. \ref{fig7}, three prominent peaks are located in the polarization of the H$\alpha$ line. To
clarify the nature of the BLR and scattering region, { we transformed the parameter (vector) $-Q$ to the radio-jet direction, 
$\sim 170^\circ$ ($U$ is perpendicular to the $Q$ vector direction) and}
 plotted  Q$_\lambda\times I_\lambda$  and U$_\lambda\times I_\lambda$ parameters
(upper panel
 in Fig. \ref{fig9}) and the projection of velocities along the H$\alpha$
profile on the UQ space (bottom panel). In Fig. \ref{fig9}  { we divided the points according the velocities, first of all from the
bumps which can be seen in Fig. \ref{fig7}, and taking that velocities from -1000 km ${\rm s^{-1}}$ to
+2000 km ${\rm s^{-1}}$  correspond to
the velocities of the Keplerian motion in the disc (denoted as  open and full circles).}
 {It is very interesting} that on the {transformed} UQ surface, 
 there is a clear separation between points {(velocities)} seen in different bumps.
 { The Keplerian disc space is projected nearly on a  circular surface, where one
axis is nearly parallel to {  the} radio-jet direction { (arrow on the plot)} and approximately separated
the points with receding (full circles) and approaching (open circles) velocities. The space
of the Keplerian disc is with smaller velocities (between -1000 km ${\rm s^{-1}}$ and +2000 km ${\rm s^{-1}}$), 
while the components A and C are with higher velocities and they are well separated. As we noted above these two components
corresponds to the blue peak and red shoulder in the H$\alpha$ line profile, which are expected  in a line emitted from a 
relativistic disc, i.e.
in orbits closer to the central black hole (effects of the Doppler boosting and gravitational redshift). 
It can be seen in Fig. \ref{fig9} that 
the component C has points that are located
in the UQ space with $Q<0$, similar as receding part of Keplerian disc (full circles), while A has points with 
$Q>0$ as the points from approaching side of the Keplerian disc. These may indicate that the components A and C are coming from the 
non-Keplerian part of the disc (closer to the central black hole). On the other hand, the component B, that cannot be seen in the 
H$\alpha$ broad line profile, has points that nearly follow $+Q$ axis that may indicate
high-speed outflow (around -6000 km ${\rm s^{-1}}$), in the inner part of the BLR\footnote{ The negative velocities in 
this polarized component indicate an amount of approaching gas, that may be outflowing material that has the near radio-jet direction}. 
}

\subsection{Central part of Mrk 6: BLR gas motion and black hole mass}

\subsubsection{Observational evidence of the Keplerian motion in the BLR}

Assuming that the $\varphi$ vs. velocity dependence is caused by only the velocity field in the BLR \citep[][]{sm05},
we are able to explore the kinematics of the emission gas in the BLR.
 We should mention here, that in the estimation of
an AGN black hole mass, an assumption of the Keplerian motion in the BLR is {\it a-priori} accepted
\citep{pe13}, but there is a few  observational indication that this
motion is dominant in the BLR.

\begin{figure}
\centering
\includegraphics[width=8cm]{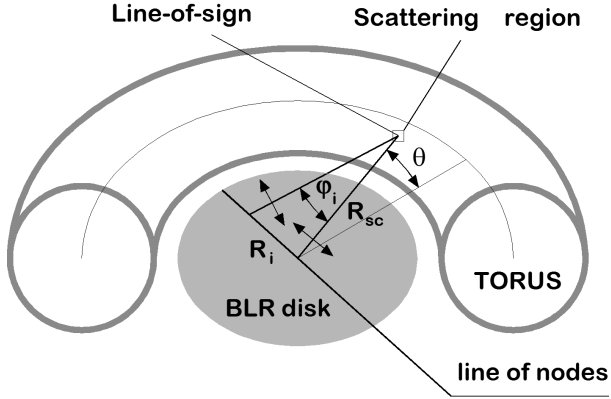}
\caption{The scheme of the disc and scattering region in the inner part of the torus} \label{fig10a}
\end{figure}

If we assume that a predominant Keplerian motion is in the BLR of Mrk 6, then observed velocities $V_i$, in each part of the
rotating disc, are depending only from the  distance $R_i$ from the
central black hole (with mass $M_{BH}$). { An illustration of the equatorial scattering on the torus
of the emission from the  Keplerian disc is given in
Fig. \ref{fig10a}.} The projected velocity in the plane of the scattering region is:
$$V_i=V^{rot}_i\cos(\theta)=\sqrt{GM_{BH}\over R_i}\cos(\theta), \eqno(1)$$
where $G$ is the gravitational constant and $\theta$ is the angle between the disc and polarization
plane {(see Fig. \ref{fig10a})}. In the case of the equatorial polarization, $R_i$ can be connected with the corresponding
polarization angle {(see Fig.\ref{fig10a})} :
$$R_i=R_{sc}\cdot \tan(\varphi_i), \eqno(2)$$
where $R_{sc}$ is the distance from the center of the disc to the scattering region.

{ Taking into account the polarization angle of different parts of the disc ($\varphi_i$, see Fig. \ref{fig10a})},
Eq. 1 can be rewritten as:

$$\log({V_i\over c})=a-b\cdot log(\tan(\varphi_i)), \eqno(3)$$
where $c$ is the velocity of light, the constant $a$ is
$$a=0.5\log\bigl({{GM_{BH} \cos^2(\theta)}\over{c^2R_{sc}}}\bigr). \eqno(4)$$
In the case of the Keplerian motion $b\approx0.5$.

\begin{figure}
\centering
\includegraphics[width=8cm]{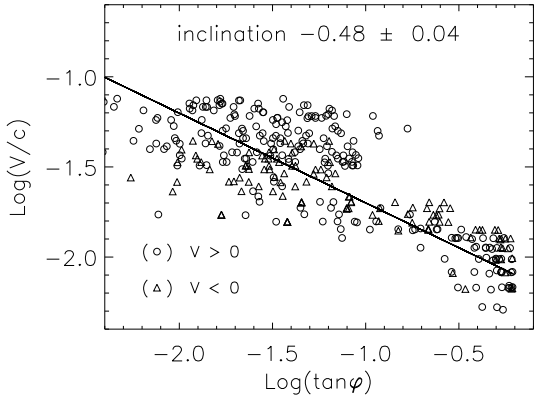}
\caption{The velocity as a function of the polarization angle across the H$\alpha$ line profile.
The zero velocity scale is taken {  to be} +1000 km ${\rm s^{-1}}$ and receding velocities are denoted with
circles and approaching with triangles. The solid line represents the best fit.} \label{fig10}
\end{figure}

The best fitting {  of} $\log(V)$ vs. $\log(\tan(\varphi))$, shown in Fig. \ref{fig10}
{  with} the solid line, gives  $b=0.48\pm0.04$ (that is practically  $b\approx0.5$).
This is the evidence that the Keplerian motion is predominant one in the
BLR of Mrk 6.

\subsubsection{Estimation of the black hole mass}

There are several methods for the super-massive black hole mass estimation
\citep[see in more details][and references therein]{pe13},
but spectro-polarometric observations have not be considered {  so far} for this task. However,
the shape of the polarization angle as a function of velocities (see Eq. 3) can also be used, at least, for
rough mass estimation.  From Eq. 4, the mass of black hole can
be calculated as:

$$M_{BH-kep}=10^{2a}{c^2R_{sc}\over{G\cdot \cos^2(\theta)}}=
  1.78\cdot10^{2a+10}{R_{sc}\over{cos^2(\theta)}}M_{\odot}, \eqno(6)$$
where $R_{sc}$ is in light days.

{ From the best fit we found
 $a=-2.19\pm0.21$}, that is used for estimation of the central black hole.
As it can be seen the mass {  is} depending on
the distance of the scattering region ($R_{sc}$) from the center of the disc.
Assuming that the scattering region is located in the inner part of
the torus, one can use the  dimension of the Mrk 6 inner radius of the  torus as $R_{sc}\approx$0.18$\pm$0.05 pc
or $R_{sc}\approx$220$\pm$60 light days \citep[estimated using radiation at 2.2 microns, see][]{kis11}.
Taking $\theta=0$ (the scattering region is  in the same plane as the disc),
we obtained $M_{BH-kep}=1.53\times 10^8M_\odot$. In the general case, the angle $\theta\not=0$ since
the scattering region is in the torus with a thickness, therefore, we estimated a low limit of the BH mass.
The estimated mass is {  in} a good agreement with ones given by \cite{gr12} and \cite{Do12},
who obtained  $M_{BH}\sim$ 1.3$\cdot10^8M_{\odot}$ and  1.8$\cdot10^8M_{\odot}$,respectively.

\section{Discussion}

Mrk 6 is an AGN with broad lines, i.e. belongs to type 1 objects. In those objects, the line of sight of
an observer has a direct view of the bright nuclear region that is inside of the torus. This allows that
different nuclear polarization in the continuum components may dominate.

In type 1 objects the optical continuum is often polarized at
P$\sim$1\% level, with $\varphi$ that is  parallel to the jet axis \cite[see e.g.][]{be90}. The observed polarization
in Mrk 6 very well {  fits}  this general picture, since the level of the continuum polarization is around
$\sim 0.5$\%, and as it can be seen  in Fig. \ref{fig5}, the polarization angle, in non-active state of the AGN.
follows the radio jet direction.  The variation observed in the polarization of the continuum (p and $\varphi$) is
caused by additional polarization effects. The change in the polarization parameters is connected with the variation in brightness of the
continuum (see Figs. \ref{fig3} and \ref{fig5}), and it may be connected with some changes in the jet, since there is
{  a change}, not only
in the percentage of polarization, but also in the polarization angle ($\sim20^\circ-30^\circ$).
 In a such scenario, the polarization in the
continuum has two components, {  the first one, that is} dominant and expected component parallel to the
jet, that does not show strong changes. The second component might be connected with the jet, or outflowing
material that is very compact ($\sim2$ light days)
and produce variability in the polarization parameters. { However, looking at SED of Mrk 6 (using NED database) and
comparing with SED of the generic blazar spectrum \citep[see][]{sa89} it seems that extrapolated
radio flux is too low to explain the optical polarization. On the other hand we cannot exclude the possibility that this (variable)
fraction of the continuum is coming from some kind of outflowing material, or that scattering of the continuum light is partly occurring
on material in jet-like structure.}
Note here that the existence of large and complex
radio structures in Mrk 6 may be explained as a result of an
episodically powered precessing jet that changes its orientation \citep[][]{Kh06}, and such precession can
produce the variation in the polarization parameters.

The broad lines in type 1 AGNs are often polarized,  but at lower $P$ and at different polarization angle
than the continuum \citep[see; e.g.][]{sm04}. Also, the  H$\alpha$ line polarization is following this rule,
but, as it can be seen in Fig. \ref{fig7}, there are three peaks indicating more complex broad line emitting
region. It seems that the BLR kinematics is complex,
but the Keplerian motion of the emission gas in a rotating disc is predominant.

{  Basically,}  the polarization in the Mrk 6 H$\alpha$ line follows the trend seen in type 1 object \citep[][]{sm05},
that implies that the scattering region is equatorial in shape.
The parallel polarization  observed in Mrk 6 ($\varphi\sim 180^\circ$), quite possibly indicates that the
scattering region is in a flattened/equatorial optically-thin geometry having its symmetry axis
along the jet direction. The shift in the zero of the H$\alpha$ polarization angle (with respect to the continuum
polarization, see Fig. \ref{fig7}) of +1000 km ${\rm s^{-1}}$ may be caused by two effects: 1) the outflowing scattering region,
and 2) the inflowing
disc due to the accreating velocity to the center of the central black hole.
Both of these effects are able to shift the zero line polarization angle to the red velocities.

The most interesting result is that the function of $\varphi$ vs. velocity is following the shape expected in the case of
the Keplerian motion in the disc, and  it is a strong evidence of the Keplerian motion in the BLR of Mrk 6.
Using this fact we estimated the low limit of the black hole mass, and this estimate is in a good agreement
with ones obtained from the reverberation {  mapping}.

\section{Conclusion}

Here we presented spectro-polarization observations of Mrk 6, obtained in more than 2-years period.
Also, we observed and estimated contribution of the ISM polarization to the observed Mrk 6 polarization.
We measured polarization parameters for the continuum at 5100 \AA\ (rest wavelength) and H$\alpha$ line
and explore the lag between the unpolarized and polarized flux. On the basis of our investigation we
can outline following conclusions:

i) The ISM polarization has a significant contribution to the observed Mrk 6 polarization. Therefore,
it is important, not only in this AGN, but in general {  to} observe and take into account the ISM polarization. It should be considered that the ISM can affect not only the
 {  degree}  of polarization, but also the polarization angle.

ii) We found that the variability in the polarized continuum follows the variability in unpolarized one, and
that the lag is $\sim 2$ days, that is ten times smaller than the estimated BLR dimension of $\sim20$ days.
This implies that the scattering region of the continuum is complex, and it seems to have a part where the
vector of polarization is with the angle of 180$^\circ$,
and {  another}  scattering region that contributes to the variation {  which}  my be some kind of outflowing scattering material.

iii) During the monitoring period, the H$\alpha$ line profile, as well as H$\alpha$ polarization parameters
 {  have not changed}  significantly.
The polarization  { shape} (as a function of wavelengths) has three bumps, { where one (the most intensive and blue-shifted)
probably is coming from
the outflow in the inner part of the BLR. On the other hand, rest
 two bumps (shifted to the blue and red), which correspond to the blue peak and red shoulder in the broad line
 profile, are probably coming from relativistic part of the accretion disc.}

iv) Comparing our observations with {  those}  performed by \cite{sm02}, we found that the line 
profile of the H$\alpha$ and its polarization
parameters are significantly changed in a long time period. The polarization angle as a function 
of velocity across the line profile save the same trend, and the zero scale of the polarization 
in angle shows a redshift of +1000 km ${\rm s^{-1}}$, that may be caused by two effects: inflowing
accreating gas in the disc, and outflowing scattering region.

v) We give an observational evidence for the Keplerian motion in the BLR of Mrk 6. 
Using spectro-polarimetric observations we roughly estimated the black hole mas as 
$\sim1.53\times10^8M_\odot$, that is a low limit of mass. However our estimate is in a 
good agreement with ones obtained by reverberation ($M_{BH}\sim1.3-1.8\times10^8M_\odot$).

Here we give {  the}  observational facts with a simple calculation to give very rough estimates,
for a more sophisticated explanation of the observational data one should apply full theory of polarization
that we postpone to the following paper (Afanasiev et al. 2012, in preparation).

\section*{Acknowledgments}
The results of observations were obtained with the 6-m BTA telescope of the Special Astrophysical
Observatory Academy of Sciences, operating with the financial support of the Ministry of Education
and Science of Russian Federation (state contracts no. 16.552.11.7028, 16.518.11.7073).The authors also express appreciation to the
Large Telescope Program Committee of the RAS for the possibility of implementing the program
of Spectropolarimetric observations at the BTA.
This work was supported of the  Russian Foundation for Basic Research (project N12-02-00857)
and the Ministry of Education and Science (Republic of
Serbia) through the project Astrophysical Spectroscopy of
Extragalactic Objects (176001). {
 We would like to thank to Prof. Makoto Kishimoto for very useful comments and suggestions.
 L.\v C. Popovi\'c thanks to the COST Action MP1104 'Polarization as a tool to study the Solar System and beyond' for support.}

\end{document}